\documentstyle[12pt]{article}
\renewcommand{\theequation}{\arabic{equation}}
\newcommand{\EQ}{\begin{equation}}
\newcommand{\EN}{\end{equation}}

\newcommand{\bear}{\begin{eqnarray}}
\newcommand{\ear}{\end{eqnarray}}

\begin{document}

\topmargin 0pt
\oddsidemargin 5mm
\newcommand{\NP}[1]{Nucl.\ Phys.\ {\bf #1}}
\newcommand{\PL}[1]{Phys.\ Lett.\ {\bf #1}}
\newcommand{\NC}[1]{Nuovo Cimento {\bf #1}}
\newcommand{\CMP}[1]{Comm.\ Math.\ Phys.\ {\bf #1}}
\newcommand{\PR}[1]{Phys.\ Rev.\ {\bf #1}}
\newcommand{\PRL}[1]{Phys.\ Rev.\ Lett.\ {\bf #1}}
\newcommand{\MPL}[1]{Mod.\ Phys.\ Lett.\ {\bf #1}}
\newcommand{\JETP}[1]{Sov.\ Phys.\ JETP {\bf #1}}
\newcommand{\TMP}[1]{Teor.\ Mat.\ Fiz.\ {\bf #1}}

\renewcommand{\thefootnote}{\fnsymbol{footnote}}

\newpage
\setcounter{page}{0}
\begin{titlepage}
\begin{flushright}
UFSCARF-TH-94-4
\end{flushright}
\vspace{0.5cm}
\begin{center}
{\large  On the parafermionic scattering of an integrable Heisenberg model with
impurity}\\
\vspace{1cm}
\vspace{1cm}
{\large  M.J.  Martins } \\
\vspace{1cm}
{\em Universidade Federal de S\~ao Carlos\\
Departamento de F\'isica \\
C.P. 676, 13560~~S\~ao Carlos, Brasil}\\
\end{center}
\vspace{1.2cm}

\begin{abstract}
We have studied the massive parafermionic
sector of an integrable spin-$s$ chain with an impurity of
spin-$s^{'}$ in presence of a magnetic
field. The effect of the impurity is encoded in a computable
parafermionic impurity scattering amplitude and
its contribution to the finite size properties of the
free energy is discussed. As an example we compute the non-integer degeneracy
of the ground state.
For the simplest model of $s=1$ and $s^{'}=1/2$, its interpretation
in terms of conformally invariant boundary conditions is presented.
\end{abstract}
\vspace{.2cm}
\vspace{.2cm}
\centerline{February 1994}
\end{titlepage}

\renewcommand{\thefootnote}{\arabic{footnote}}
\setcounter{footnote}{0}

\newpage
\section{Introduction}
Recently there has been renewed interest on the study of physical properties of
spin chains in presence
of an impurity \cite{AF,AF1}. An impurity may be considered as an extra
magnetic spin interacting with
magnetic ions of a homogeneous chain \cite{AF1,AF2}. Although an imputity is
not
supposed to change the basic
bulk properties, one should expect that it may affect in a non-trivial way the
surface behaviour. An example
of a system possessing this feature is an integrable spin-$s$ Heisenberg chain
with an impurity
of spin-$s^{'}$ \cite{HO,SO,SO1}. The bulk spectrum is gapless and the
underlying conformal field theory is that
of the $SU(2)$ Wess-Zumino-Witten-Novikov model at level $j=2s$ \cite{AF3,WE}.
For $s \leq s^{'}$ this impurity model
behaves as a spin chain with an extra site and a decoupled effective spin
$s^{'}-s$ \cite{SO1,AF2}. However,
when $s > s^{'}$, the extra spin degrees of freedom induce a non-trivial
critical behaviour in the impurity
contribution to the thermodynamic properties \cite{SO1}. In fact, this surface
behaviour resembles much that of a
spin $s^{'}$ impurity interacting with $k(k=2s)$ degenerated bands of free
electrons, namely the one impurity
Kondo effect \cite{LU,LU1}.

The purpose of this paper is to study the effect of the impurity $s^{'}$ on the
$Z(2s)$ parafermionic sector of the
bulk system in the ``overscreened'' regime $s > s^{'}$. Our strategy is to
analyze the integrable impurity
model in presence of a small magnetic field. This is motived by the known fact
that a magnetic field is able
to break the $SU(2)_{2s}$ spectrum (of the usual integrable spin-$s$ Heisenberg
model \cite{BA}) in a massless
bosonic field and a massive $Z(2s)$ parafermionic field theory \cite{TV}. In
this sense, similar approach is used
in the impurity system in order to isolate the parafermionic degrees of
freedom. In the interesting regime
of $s <s^{'}$ we shall find that the main effect of the impurity is to
introduce an extra scattering amplitude
between the ``impurity'' and the $2s-1$ particles composing the $Z(2s)$ massive
spectrum. Taking into account
the impurity scattering, we make use of the thermodynamic Bethe ansatz approach
\cite{YY,ZA} in order to study
the impurity finite-size effects of the free energy in the ultraviolet limit.
Remarkably, we find that the
parafermionic impurity scattering alone is enough to describe the non-integer
degeneracy of the ground state of the
impurity Heisenberg chain for $s < s^{'}$.

This paper is organized as follows. In sect.2 the parafermionic degrees of
freedom are isolated, the impurity
scattering amplitude is computed and the corresponding thermodynamics is
formulated. In sect.3 we discuss
the finite size properties of the impurity free energy in the high temperature
regime. Sect. 4 is devoted
to some comments on our results. In particular, for the simplest case of $s=1$,
the role of conformal boundary
conditions \cite{CA,CA1} is discussed in the context of the degeneracy of the
ground state \cite{LU2}. In
appendices A and B we summarize useful relations concerning the finite size
effects of the impurity free
energy.

\section{ The impurity parafermionic scattering and its thermodynamics}

We start by briefly reviewing the basic Bethe ansatz results of the integrable
spin-$s$ chain
in presence of a spin-$s^{'}$ impurity. The model was first solved in ref.
\cite{HO} for $s=1/2$ and
afterwords was generalized to include all spin values \cite{SO,SO1}. Due to
integrability,
the associated Hamiltonian consisting of a fine-tuned interaction between $L$
spins-$s$ and one impurity
of spin-$s^{'}$ can be diagonalized by the Bethe ansatz approach
\cite{HO,SO,SO1}. As usual, in
the thermodynamic limit, the Bethe ansatz equations can be written in terms of
the
densities of particles($\sigma_n(\lambda))$ and holes($\sigma_n(\lambda))$ of a
given configuration of
$n$-string type for the rapidities $\lambda$. In our case these equations are
\EQ
2 \pi \tilde{\sigma}_n(\lambda)= \psi_{n,s}(\lambda)
+\frac{\psi_{n,s^{'}}(\lambda)}{L} -2\pi
\sum_{m=1}^{\infty}[A_{n,m}*\sigma_{m}](\lambda)
\EN
where the symbol $[f*g](x)$ stands for the convolution $\frac{1}{2 \pi}
\int_{-\infty}^{+\infty}
f(y)g(x-y)dy$. The Fourier transform of functions $A_{n,m}(\lambda)$ and
$\psi_{n,m/2}(\lambda)$ have
the following simple expressions
\begin{eqnarray}
A_{n,m}(\omega)= \coth(|\omega|/2) \left [ e^{-|n-m||\omega|/2}
-e^{-(n+m)|\omega|/2} \right ] \\
\psi_{n,m/2}(\omega) = A_{n,m}(\omega) p(\omega)
\end{eqnarray}
where $p(\omega)=\frac{1}{2 \cosh(\omega/2)}$, and the Fourier component of
$f(x)$ is defined by
$f(\omega)= \frac{1}{2 \pi} \int_{-\infty}^{+\infty}f(x) e^{-i\omega x}dx$.

The thermodynamic properties are encoded in terms of the excitation energies
$\epsilon_n(\lambda)= T \ln[\tilde{\sigma}_n(\lambda)/\sigma_n(\lambda)];
n=1,2, \cdots, \infty$. At
temperature $T$ and in a magnetic field $H$, those energies satisfy the
following thermodynamic Bethe ansatz
equations \cite{SO1}
\EQ
T \ln(1 +e^{\epsilon_n(\lambda)/T)})= -\psi_{n,s}(\lambda) +nH +
T \sum_{m=1}^{\infty} \left [ A_{n,m}*\ln(1+e^{-\epsilon_m/T}) \right ]
(\lambda)
\EN

Here we are first interested on the $ T \rightarrow 0 $ behaviour of functions
$\epsilon_n(\lambda)$
for a small magnetic field $H <<1$. Eq.(4) is the same kind of that appearing
in the thermodynamics of the
integrable spin-$s$ chain \cite{BA}, and the behaviour of $\epsilon_n(\lambda)$
has been already considered
by Tsvelick \cite{TV}. In this regime, the mode $n=2s$ is considered as the
ground state of the system and
the low-lying massive excitation corresponds to the lightest mass gaps
$\epsilon_n(\lambda); n=1,2, \cdots, 2s-1$.
The remaining modes scale directly with the magnetic field, and are the
heaviest decoupled massive excitations of the
system. Taking into account this discussion, our main task now is to isolate
the low-lying modes in Eq.(1)
in terms of the ground state excitation $\sigma_{2s}(\lambda)$. First we invert
Eq.(1) for functions
$\tilde{\sigma}_n(\lambda)$ and $\sigma_n(\lambda)$, obtaining the following
expression
\EQ
\sigma_n(\lambda) +\tilde{\sigma}_n(\lambda) = p*[ \tilde{\sigma}_{n+1}
+\tilde{\sigma}_{n-1}(\lambda)](\lambda) +\frac{p(\lambda)}{ 2\pi}(
\delta_{n,2S}
+\frac{\delta_{n,2S^{'}}}{L})
\EN

Considering that the physical modes are $n=1,2, \cdots, 2s-1$, it is easy to
notice from Eq.(5) that the impurity
term $\delta_{n,s^{'}}/L$ will affect the solution only when $s >s^{'}$. In
this case we solve Eq.(5) iteratively, and
we find the following result for $\tilde{\sigma}_n(\lambda)$
\EQ
\tilde{\sigma}_n(\lambda) + \sum_{m=1}^{2s-1} [B_{n,m}* \sigma_m](\lambda)=
[B_{n,2s}*p* \tilde{\sigma}_{2s}](\lambda) +\frac{1}{2 \pi L}
[B_{n,2s^{'}}*p](\lambda);~n=1,2, \cdots, 2S-1
\EN
where the matrix elements $B_{n.m}$ are given by
\EQ
B_{n,m}^{-1}(\lambda)= \delta_{n,m} -p(\lambda)( \delta_{n,m+1}
+\delta_{n,m-1})~~~n,m=1,2, \cdots 2s-1
\EN

Following  Tvelisck's approach \cite{TV} one can show that (for $H<<1$) the
triple convolution term of Eq.(6)
is proportional to the mass term induced in the system, namely
\EQ
[B_{n,2s}*\sigma_{2s}*p](\lambda) \sim m_0 \sin(\pi n/2s) \cosh(\pi \lambda/s)
\EN
where $m_0=\frac{1}{s}\int_{-\ln(H)/\pi}^{\infty} e^{-\lambda \pi/2s}
\sigma_{2s}(\lambda) d \lambda$, is the
bare scale depending on the magnetic field $H$. Introducing
the relativistic rapidities $\theta=\pi \lambda/s$
and substituting Eq.(8) into Eq.(6) we finally have
\EQ
\tilde{\sigma}_n(\theta) + \sum_{m=1}^{2s-1} [\tilde{B}_{n,m}*
\sigma_m](\theta)=
m_0 \sin(n \pi/2s) \cosh(\theta) +\frac{1}{2 \pi L}
[\tilde{B}_{n,2s^{'}}*\tilde{p}](\theta);~n=1,\cdots,2s-1
\EN
where $\tilde{p}(\omega)=\frac{1}{2\cosh(\pi \omega/2s)}$ and
$\tilde{B}_{n,m}^{-1}(\theta)=
\delta_{n,m} -\tilde{p}(\theta)( \delta_{n,m+1} +\delta_{n,m-1})$.

The next step is to find the set of discrete Bethe ansatz equations which shall
produce Eq.(9) in the
thermodynamic limit. The terms involving the densities
$\tilde{\sigma}_n(\theta)$, $\sigma_n(\theta)$ and
the mass are typical of those appearing in the Bethe ansatz equations of
diagonal factorizable
$S$-matrices with $2s-1$ particles. The impurity contribution is due the last
term of Eq.(9) and it
can be interpreted as coming from the scattering between the $2s-1$ particles
and one
static impurity \footnote{ This is in accordance with the fact that the last
term of Eq.(9) depends only
on the particle rapidities $\theta$.}. Thus, let us consider a system of $N$
relativistic particles
and one impurity scattering through diagonal amplitudes. We denote by
$S_{ab}(\theta)$ and $S_{aI}(\theta)$
the amplitudes of the particle-particle and the particle-impurity scattering,
respectively. As
usual \cite{ZA} this system can be quantized on a box of size $L$ by the
following Bethe ansatz
equations
\EQ
e^{iLm_a \sinh(\theta_i^a)} \prod_{b=1}^{N} \prod_{j \neq i}
S_{ab}(\theta_i^a-\theta_j^b)
S_{aI}(\theta_i^a)=1
\EN

Taking the thermodynamic limit of Eq.(10) and using definitions
$\phi_{ab}(\theta)= -i\frac{d}{d \theta} \ln[S_{ab}(\theta)]$ and
$\phi_{aI}(\theta)= -i\frac{d}{d \theta} \ln[S_{aI}(\theta)]$ we obtain a set
of continuous equations
for the densities $\tilde{\sigma}_{a}(\theta)$ and $\sigma_a(\theta)$ given by
\EQ
2 \pi(\tilde{\sigma}_a(\theta) +\sigma_a(\theta))= m_a\cosh(\theta)
+\frac{\phi_{aI}(\theta)}{L}
+ 2 \pi \sum_{b=1}^{N} [\phi_{ab}* \sigma_b](\theta)
\EN

By comparing Eqs.(9,11) we are able to identify the mass ratios
$\frac{m_a}{m_1}= \sin(a \pi/2s)/\sin(\pi/2s)$ and the amplitude
$S_{ab}(\theta)$ as those appearing
on the particle content of the $Z(2s)$ scattering \cite{TV}. On the other hand,
the impurity amplitude
$S_{aI}(\theta)$ has the following expression
\EQ
S_{aI}(\theta)= \prod_{a=1}^{min(a,2s^{'})} F_{|2s^{'}-a|+2l-1}(\theta),~~~~
F_n(\theta)= \frac{\sinh\frac{1}{2}(\theta-i\pi n/2s)}
{\sinh\frac{1}{2}(\theta+i\pi n/2s)}
\EN

It is important to observe that the impurity scattering has no pole on the
physical strip
$0 <Im(\theta) <\pi$. This means that the impurity can not induce  extra
excited states( impurities)
on the system. This fact was an implicit assumption in writing the discrete
bethe
ansatz equations of form (10), and in this sense
our approach is consistent.

The rest of this section is concerned with the analysis of the impurity
contribution to the thermodynamic
properties of the system. We recall that the finite size effects of the ground
state energy $E(R)$ on
the cylinder of radius $R$ is related to the properties of the free energy at
finite temperature $T=1/R$
\cite{AF3,CA3,ZA}. Following the standard approach \cite{YY,ZA} the temperature
is encoded through
the minimization of the free energy in terms of densities of particles and
holes constrained
by the ``impurity'' Bethe ansatz equation (11). At this point it is convenient
to decompose
the total free energy in terms of its bulk ($F_b$) part and the impurity
($F_{I}$) contribution as
\EQ
F= LF_b +F_I
\EN

After the minimization we find
\EQ
F_b = -\frac{1}{2 \pi R} \sum_{a=1}^{N} m_a\int_{-\infty}^{+\infty}
\cosh(\theta) \ln(1+e^{-\epsilon_a(\theta)})
\EN
\EQ
F_I = -\frac{1}{2 \pi R} \sum_{a=1}^{N} \int_{-\infty}^{+\infty}
\phi_{aI}(\theta)
\ln(1+e^{-\epsilon_a(\theta)})
\EN
where the excitation energy $\epsilon_a(\theta)$ satisfies the following
thermodynamic Bethe ansatz equations
\EQ
\epsilon_a(\theta) +\sum_{b=1}^{N} [\phi_{a,b}* \ln(1
+e^{-\epsilon_b})](\theta) = m_a R \cosh(\theta)
\EN

As it would be expected the thermodynamic Bethe ansatz equation is only
dependent on the bulk
scattering quantities, namely $\phi_{ab}(\theta)$ and $m_a$. However,
concerning with the free
energy( besides the usual bulk term), we have an extra impurity part which is
strongly dependent
on the scattering amplitudes between the particles and the impurity. In the
next section we are going
to study the finite size effects of the impurity free energy (15).

\section{The impurity finite size properties}
Here we are interested in the study of the finite size effects of the impurity
free energy. From the usual
scale arguments, one would expect that function $f_I$=$RF_I(R)$ is scaling
invariant. In other words,
all the length dependence enters as a function of the scale invariant r=mR
\footnote{m is considered
to be the lowest mass gap in the system.} parameter. Analogously to the bulk
case \cite{AF3,CA3}, this
function posses a universal behaviour in the $r \rightarrow 0$ limit. In this
regime, we expect
to find the following behaviour
\EQ
f_I(r)= \varepsilon_I r -\ln(g) +O(r^y) ,~~~ r \rightarrow 0
\EN
where g is interpreted as the ground state degeneracy and it is expected to
behave as a universal number \cite{LU2}.
The first term of Eq.(17) can be thought as a consequence of extending the
hyperscaling hypothesis to
the surface properties, namely $f_I(r) \sim m$ ( in our case the correlation
length $\xi \sim 1/m$). The
exponent $y$ is connected to the leading finite size corrections away from the
critical point. For instance,
if the perturbating surface field has a holomorphic conformal dimension
$\Delta$, standard arguments of
perturbation theory \footnote{ It is assumed that the one point function of the
perturbating operator on the
surface
is not null \cite{CA,CA1}.}
would predict that $y=2(1-\Delta)$.

Now we turn to the computation of the values of the constants $\varepsilon_I$
and $g$ of our model. The calculation of $\varepsilon_I$ is a bit technical and
for this reason is summarized in appendix A. Its value is given by
\EQ
\varepsilon_I = \frac{\sin(\pi s^{'}/s)}{\sin(\pi/s)}
\EN

The universal number $g$, however, can be estimated from the expression (15) as
follows. Due to the behaviour
of functions $\phi_{aI}(\theta)  \sim e^{-\theta}$ on the $\theta \rightarrow$
limit, the fluctuations
on the $r \rightarrow 0$ regime of the impurity free energy are dominated by
the strictly $r=0$ behaviour
of the excitation energies $\epsilon_a$. Therefore, by an explicit integration
of Eq.(15) we find that
\EQ
\ln(g) =\sum_{a=1}^{2s-1} [2-l]_{a,2s^{'}}^{-1} \ln(1+e^{-\tilde{\epsilon}_a})
\EN
where $l$ is incident matrix of a $A_{2s-1}$ Lie algebra. The constants
$\tilde{\epsilon}_a$ exhibit
the $r=0$ behaviour of the bulk thermodynamic equations (16), and satisfy the
following relation
\EQ
\tilde{\epsilon}_a -\sum_{b=1}^{2s-1}(l[2-l]^-1)_{ab}
\ln(1+e^{-\tilde{\epsilon}_b}) =0
\EN

By comparing these two last expressions it is easy to find the following
relation
\EQ
\ln(g)= \frac{1}{2}  [\tilde{\epsilon}_{2s^{'}}
+\ln(1+e^{-\tilde{\epsilon}_{2s^{'}}})]
\EN
and by using the following value of $e^{\tilde{\epsilon}_{2s^{'}}}$
\cite{MA,KA}
\EQ
e^{\tilde{\epsilon}_{2s^{'}}}= \frac{\sin^2 [ (2s^{'}+1)\pi)/(2s+2)]}
{\sin^2[\pi/(2s+2)]} -1
\EN
we finally find that
\EQ
g= \frac{\sin [ \pi(2s^{'}+1)/(2s+2)]}{\sin[\pi/(2s+2)]}
\EN

Remarkably enough, this is precisely the value obtained in ref. \cite{SO1} for
the zero temperature entropy
of the integrable spin-$s$ chain with an impurity of spin-$s^{'}$. At this
point we recall that the
bulk spectrum of this model is given by a certain combination of $c=1$ bosonic
field theory and a $Z(2s)$
parafermions \cite{AF3,WE}. Thus, this leads to the conclusion that the
non-trivial behaviour of the degeneracy
of the ground state is due only to the parafermionic sector.

Let us  analize the next to the leading corrections determined by the exponent
$y$. In the case of the
model with $s=1$, the finite size effects can be exactly calculated as shown in
appendix B. We conclude
that the most singular part of the impurity free energy has a logarithmic
singularity. This is a typical
feature of a marginally irrelevant surface field scaling with $y=1$. For the
other theories, this exponent
can be estimated by numerically solving the system of Eqs.(15,16). For each
distinct points $r_1$ and $r_2$
the exponent $y$ can be computed by extrapolating the following sequence
\EQ
y(r_1,r_2)= \ln\left[(f_I(r) -\epsilon_I r_1 +\ln(g))/(f_I(r_2) -\epsilon_I r_2
 +\ln(g))
\right ]/\ln(r_1/r_2)
\EN

In table 1 we present such estimatives for several values of the spins $s$ and
$s^{'}$. The scaling
dimension $y$ depends only of the bulk theory(as it is expected to be)
and our numerical data is consistent with the value $y=2s/(s+1)$. Hence, we
conclude that the surface
perturbating field has the same holomorphic conformal dimension of the operator
which deformed the $Z(2s)$
parafermionic theory, namely the energy operator with $\Delta=1/(s+1)$.
However, the coefficient proportional to $r^{y}$ is spin-$s^{'}$ dependent. By
defining $f_I(r)- \varepsilon_I r +\ln(g) = r^{y} $ up to
order $O(r^{y})$, we are able to estimate the constants $b(s,s^{'})$. In table
2 we show some of these
estimatives. We notice that due to the $Z(2s)$ symmetry one can derive the
relation $b(s,s^{'})=b(s,s-s^{'})$.
In general, one can use these numbers in order to compare with those coming
from the perturbation
theory around a conformally
invariant fixed point with boundaries \cite{CA,CA1}. In the next section we
comment on this point for
the simplest case of $s=1$.

\section{Comments}
To conclude this paper we would like to make the following remarks. It seems
important to recall that,
recently, factorizable $S$-matrices have been studied in presence of boundary
conditions \cite{KE,ZAM}. Roughly,
in this approach, the analogous of our impurity $S$-matrix is the amplitude
describing the
bulk particle scattering on the boundary state. In general, for the $Z(N)$
diagonal scattering, one can
verify that our impurity amplitudes satisfy quite different properties of those
obtained for the boundary
scattering\cite{KE,ZAM,JA}. Interesting enough, one exception is the case of
the simplest impurity theory of $s=1$ and $s=1/2$.
Using Eq.(12) we obtain $S_{1I}=-i\tanh\frac{1}{2}(\theta -i \pi/2) $, which is
precisely the boundary
amplitude appearing in the Ising field theory with fixed
boundary conditions \cite{ZAM}. In this sense we believe
that it is interesting to connect some of our finite-size results for this
model to the theory of the
conformally  invariant boundary conditions developed by Cardy \cite{CA}. One
possible way is to compute the
degeneracy of the ground state and compare it with the formula (22). The
degeneracy $g$ for conformally
invariant boundary condition $A,B$ on the semi-infinite cylinder is given by
\cite{LU2}
\EQ
g=\sum_{a} \eta_{AB}^a S_a^0
\EN
where $\eta_{AB}^a$ is the multiplicity of the character $\chi_a$ (entering in
the partition
function with boundaries $A,B$) and $S_a^0$ is the modular $S$-matrix project
out on the ground state. It
is possible to reproduce the degeneracy of the $s=1$ and $s^{'}=1/2$ model as
follows. The partition function
of the Ising model with free boundary conditions in terms of the fields
$I$(identity),
$\sigma$(magnetic operator) and $\epsilon$(energy operator) can be written as
\cite{CA2}
\EQ
Z_{ff}= \sum_{a} \eta_{ff}^a \chi_a = \chi_{I} + \chi_{\epsilon}
\EN

By fusing \cite{CA} this boundary condition with the magnetic field $\sigma$,
the new multiplicities are
$\tilde{\eta}^a = \sum_{k} N_{\sigma k}^a \eta_{ff}^k$,
where $N_{\sigma k}^a$ is the fusion rule coefficients
of the Ising model. The new partition function is now given by
\EQ
\tilde{Z}= \sum_{a} \tilde{\eta}^a \chi_a = 2 \chi_{\sigma}
\EN
and by using the modular $S$-matrix \cite{CA} of the Ising model and Eq.(24) we
find that $g=\sqrt{2}$, in
accordance with our formula (22) for $s=1$ and $s^{'}=1/2$. From the pioneering
paper of Cardy \cite{CA2},
we can identify it as a $Z(2)$-invariant combination of the characters of a
mixed boundary condition.

In general, one would expect that the finite size effects of an impurity
problem will renormalize
to critical points with certain boundary conditions. For instance, for the
Kondo effect,
Affleck and Ludwig \cite{LU,LU2} have been able to identify it as certain fused
free fermion boundary
condition. Remarkably, the overscreened degeneracy is the same of that found in
sect.3 if we
substitute the number of electron's bands by $2s$. In this sense, it seems
plausible to believe that
our results of sect.3 (for general s) may also be view as peculiar boundary
conditions on the parafermionic
bulk system. We hope to report on this topic as well as to perform extra
perturbative checks on a future
work.

\section*{Acknowledgements}
This work was partially supported by CNPq~(Brazilian agency).

\vspace{1.0cm}

\centerline{\bf Appendix A }

\vspace{0.5cm}

\setcounter{equation}{0}
\renewcommand{\theequation}{A.\arabic{equation}}
In this appendix we present the main steps leading to the surface behaviour of
Eq.(17). By
multiplying function $f_I(r)$ by $r$ and then taking the derivative we have
\EQ
\frac{d}{dr}(rf_I(r)) =-\frac{1}{2\pi} \sum_{a=1}^{N} \left [
\int_{-\infty}^{+\infty}d \theta
\phi_{aI}(\theta) \ln(1+e^{-\epsilon_a(\theta)}) d \theta -
\int_{-\infty}^{+\infty}  d \theta r \phi_{aI}(\theta) \frac{\partial}{\partial
r} \epsilon_a(\theta)
\frac{e^{-\epsilon_a(\theta)}}{1+e^{-\epsilon_a(\theta)}} \right ]
\EN

In the regime $r \rightarrow 0$ function $\epsilon_a(\theta)$ is approximately
constant in the region
$ |\theta| << \ln(2/r) $ \cite{ZA}. Considering the change of variables $\theta
\rightarrow
\theta+\ln(2/r)$,
and by observing that function $\phi_{aI}(\theta)$ behaves as
\EQ
\phi_{aI}(\theta) = f_{aI} e^{-\theta}, \theta \rightarrow \infty
\EN
we are able to rewrite Eq.(A.1) as
\EQ
\frac{d}{dr}(rf_I(r)) =-\frac{1}{\pi}r \sum_{a=1}^{N} f_{aI}
\int_{-\infty}^{+\infty} d\theta
\frac{e^{-\theta}}{1+e^{\epsilon_a(\theta)}} [ \frac{1}{r}
\frac{\partial}{\partial \theta}
\epsilon_a(\theta)+
\frac{\partial }{\partial r}\epsilon_a{\theta} ]
\EN

In this regime, the right hand side of (A.3) can be further approximate only in
terms of derivatives
on $\frac{\partial}{\partial \theta} \epsilon_a(\theta)$ \cite{ZA}, and we have
\EQ
\frac{d}{dr}(rf_I(r)) =\frac{1}{\pi}r \sum_{a=1}^{N} f_{aI} T_a
\EN
where $T_a=\int_{-\infty}^{+\infty}d \theta
e^{-\theta}\frac{\partial}{\partial \theta}\ln(1+e^{\epsilon_a(\theta)})  $.
On the other hand, from the kernel of the impurity scattering we find the
important relation
\EQ
f_{aI}= 2 \sin(\pi/2s) \frac{m_a}{m_1} \frac{m_{2s^{'}}}{m_1}
\EN
and substituting it in Eq.(A.4) we have
\EQ
\frac{d}{d r}[r f_I(r)]= \frac{2}{\pi} \frac{ m_{2s^{'}}}{m_1} \sin(\pi/2s)
\sum_{a=1}^{N}
\frac{m_a}{m_1} T_a
\EN

Finally, by using the identity $\sum_{a=1}^{N} \frac{m_a}{m_1} T_a =
\frac{\pi}{\sin(\pi/s)}$
\cite{KA}valid for the $Z(2s)$
bulk $S$-matrices and the value of $m_{2s^{'}}$, we finally obtain
\EQ
\frac{d}{d r}[r f_I(r)]= 2r \frac{ \sin(\pi s^{'}/s)}{\sin(\pi/s)}
\EN
which leads to the behaviour
\EQ
f_I(r)= r\frac{ \sin(\pi s^{'}/s)}{\sin(\pi/s)},~~~ r \rightarrow 0
\EN
\vspace{1.0cm}

\centerline{\bf Appendix B }

\vspace{0.5cm}

\setcounter{equation}{0}
\renewcommand{\theequation}{B.\arabic{equation}}
In the case of the model with $s=1$ we have that $\epsilon(\theta)=r
\cosh(\theta)$ ($S_{11}=-1$). Hence
the scaled impurity free energy is
\EQ
f_{I}(r) = -\frac{1}{\pi} \int_{0}^{+\infty}d \theta
\frac{\ln(1+e^{-r\cosh(\theta)})}{\cosh(\theta)}
\EN

The derivative of $f_{I}(r)$ is then given by  expression
\EQ
\frac{d}{d r}[f_{I}(r)]
 = \frac{1}{\pi} \int_{0}^{+\infty}\frac{d \theta}{ 1+e^{-r\cosh(\theta)}}=
\frac{1}{\pi}\sum_{k=1}^{\infty} (-1)^{k-1} K_0(kr)
\EN
where $K_0(x)$ is the modified zero order Bessel function. By using certain
sums properties of Bessel
function (\cite{KA}, see 8.526 of ref.\cite{TAB}) and by integrating back
Eq.(B.2) we find up to order of $O(r^3)$ the
following behaviour
\EQ
f_{I}(r) = -\frac{1}{2}\ln(2) -\frac{r}{2 \pi}[ C -\ln(\pi) -1] -\frac{r}{2
\pi} \ln(r) +O(r^3)
\EN
where C is the Euler constant.
\vspace{1.0cm}
\newpage
\centerline{\bf Tables }

{\bf Table 1} The extrapolated values of the exponent $y(s)$ for the models
with
$s=3/2,2,5/2,3$ and $s^{'}=1/2,1,3/2$.
\vspace{1.5cm}\\
\vspace{0.15cm}
\begin{tabular}{|l|l|l|l|l|} \hline
$s^{'}$  $ y(3/2)$  $y(2)$  $y(5/2)$  $y(3)$  \\ \hline \hline
1/2 1.20004($\pm 1$)  1.3334($\pm$ 1) 1.4287($\pm2$)  1.5001($\pm 1$)     \\
\hline
1 1.20004($\pm 1$)  1.3334($\pm$ 1) 1.4286($\pm 1$)  1.5001($\pm 1$)     \\
\hline
3/2 ~$-  1.3334($\pm$ 1) 1.4287($\pm 2$)  1.50009($\pm 2$)     \\ \hline
exact 6/5  4/3 10/7   3/2    \\ \hline
\end{tabular}\\
\vspace{2.0cm}\\
{\bf Table 2} The extrapolated values of the constants $b(s,s^{'})$ for the
models with
$s=3/2,2,5/2,3$.
\vspace{1.5cm}\\
\vspace{0.15cm}
\begin{tabular}{|l|l|l|l|l|} \hline
$s^{'}$  $ Z(3)$  $Z(4)$  $Z(5)$  $Z(6)$  \\ \hline \hline
1/2 -0.6835($\pm 2$)  -0.3624($\pm$ 1) -0.2568($\pm1$)  -0.2048($\pm 1$)     \\
\hline
1 -0.6835($\pm 2$)  -0.5436($\pm$ 1) -0.4628($\pm 1$)  -0.4101($\pm 2$)     \\
\hline
3/2 ~$-  -0.3624($\pm$ 1) -0.4628($\pm 1$)  -0.4949($\pm 1$)     \\ \hline
\end{tabular}\\

\end{document}